\documentclass[preprint,12pt]{revtex4} 
\usepackage{graphicx,amsfonts,amsbsy}
\usepackage[dvips]{epsfig,color}
\begin{document}
%
\title{Electron pairing and evidence of a BCS-BEC crossover in $d$-wave superconductors}
\author{Subrat Kumar Das$^{a,}$\footnote{Corresponding author \\ Current Address: TCMP Division, Saha Institute of Nuclear Physics, Kolkata - 700064, India.}, Ayan Khan$^{b,}$\footnote{Current Address: Department of Physics, Bilkent University, 06800 Bilkent, Turkey.}, Saurabh Basu$^{a}$}
\address{$^a$Department of Physics, Indian Institute of Technology Guwahati, Guwahati, Assam - 780039, India \\
$^b$Department of Physics, Indian Institute of Science Education and Research-Kolkata, Mohanpur Campus, West Bengal - 741252, India}
\begin{abstract}
We have demonstrated that it is possible to access a crossover scenario starting with a weak coupling (BCS) $d$-wave superconductor to a strongly coupled 
Bose-Einstein condensate (BEC) phase as the exchange interaction is tuned in a two dimensional system described by a $t$-$J$-$U$ model via numerically solving 
the Bogoliubov-de Gennes (BdG) equations. While in the extreme dilute limit, the electronic pairing phenomena is independent of the Coulomb repulsion, $U$, 
the superconductivity depends on $U$, and so does the crossover. Further, the effect of variation in the carrier density on the BCS-BEC crossover has also been investigated. 
The crossover picture is illustrated by computing the chemical potential, which when falls below the noninteracting band minimum, signals the onset of a phase with 
tightly bound, shorter pairs. As an evidence of the above feature, the Cooper pair radius is calculated which shows a significant shortening at the emergence 
of a BEC-like phase. Besides, in contrast to the previous work where a crossover was claimed only in the dilute limit, we have demonstrated it at large 
densities near half filling.
\end{abstract}
%
%
\maketitle
%
\section{Introduction}
%
The phenomena of BCS superconductivity to Bose-Einstein condensation crossover is an old one\cite{eagles, leggett, noziers}
and is interesting due to its possible connection to pseudogap observed in high-$T_c$ superconductors (HTSC)
\cite{chen2, micnas, ranninger} and also understanding the rich phase diagram of these materials.\cite{bednorz, randeria1}
Experimental data suggest that the pairing symmetry in the HTSC cuprates is predominantly of $d_{x^2-y^2}$ wave symmetry.\cite{ding, tsuei}
Unlike the conventional BCS superconductors, these compounds are characterized by extremely small pairs having a spatial extension of the order of 
one (or a few) lattice spacing(s). Randeria and coworkers shown that in two dimensional (2D) systems, a two-body bound state in vacuum is necessary for an $s$-wave ($l = 0$) pairing instability, which is not true for the higher angular momentum channels ($l > 0$).\cite{randeria2} 
The threshold for a single bound state of two electrons on the empty lattice of a 2D $t$-$J$ model are $J_c^s = 2t$ and $J_c^d = 2t/(4/\pi - 1) \simeq 7.32t$ for $s$- and $d$-wave symmetry, respectively.\cite{kagan, charu} 

The BCS superconductivity emphasizes large, overlapping Cooper pairs formed by weak attractive interaction among the electrons. 
With increasing the attractive potential, the Cooper pairs smoothly evolve into nonoverlapping composite bosons, the so called BEC 
with larger binding energy.\cite{yanase, chen}
Earlier studies mainly focused on the evolution of the coherence length of a pair of fermions at zero temperature.\cite{pistolesi, pistolesi2, casas, randeria2, randeria3} The combined effects of density and interparticle potential show that crossover is robust for all densities in case of $s$-wave pairing 
and only for low densities in case of $d$-wave pairing.\cite{hertog} 
Further, in 2D $d$-wave systems, the existence of a finite range of potential and the next-nearest-neighbor hopping drastically influence the crossover diagram.\cite{andrenacci, soares}
Experimentally measurable thermodynamical properties have also been calculated as a function of density and interaction strength 
across the crossover.\cite{duncan, pieri, singer} 
Although, the crossover phenomena is investigated from various different aspects, however a conclusive study on the BCS-BEC crossover in a d-wave SC is still pending.   
 
Coulomb repulsion enforces the no-double-occupancy constraint and has been proposed to be the key ingredient for HTSC.\cite{pwa}
Recently the $t$-$J$-$U$ ($U$ $>$ 0) model, known as the Gossamer Hamiltonian has been successfully used to describe the ground state phase diagrams of HTSC.\cite{bernevig, zhang, gan1, yuan, gan2, heiselberg, herbut}
At half-filling, the Gossamer SC state undergoes a quantum phase transition to an antiferromagnetic Mott insulator with increasing 
on-site Coulomb repulsion.\cite{bernevig, zhang, gan1, yuan} 
Near half-filling, the Mott insulator evolves into a resonating valence bond SC state\cite{gan2} as we increase the Coulomb repulsion. Further, a phase separation between antiferromagnetic and $d$-wave superfluid phases appears for $U \geq 7.3t$,\cite{heiselberg} which indicate the role of Coulomb repulsion on spin and charge density waves and stripes in cuprates. 
Apart from describing the competing orders, $t$-$J$-$U$ model is useful for a better understanding of electronic inhomogeneity and local electron density of states observed in scanning tunneling microscopy measurements.\cite{wang} 
$U\rightarrow\infty$ limit projects out all the doubly occupied sites and renders a more familiar variant of the model, normaly the $t$-$J$ model.

Considering the important role of on-site Coulomb repulsion in HTSC, we investigate its impact on the critical pairing strength for bound state formation and on the 
BCS-BEC crossover near half filling. We use the $t$-$J$-$U$ model and recalculate the threshold for two particle bound state $J_c^d \simeq 7.32t$ for 
completeness.\cite{kagan} 
This result is independent of $U$ as it should be, since $U$ has no role in a two particle intersite pairing. In fact, this result will be contrasted for pairing 
in a many-body system.  
For the many-particle case, we employ the self-consistent Bogoliubov-de Gennes (BdG) method to compute the relevant physics. We define $J_{sc}$ as the threshold  
below which $d$-wave correlations are zero, and the significance of which is illustrated later. We find that $J_{sc}$ initially increases with density and then decreases closer to half filling ($n$=1), showing a characteristic optimization behavior. We also find the signature of BCS-BEC crossover close to half filling and show that the chemical potential crosses the noninteracting band minimum with increasing the pair correlation strength. Our claim is well supported by the results obtained for the Cooper pair radius. 
It is worth while to point out that while the BdG studies on $s$-wave superconductors are fairly abundant, similar attempts on $d$-wave systems are limited. 
The problem is comprised by the fact that the attention that $d$-wave systems recieved is mostly restricted by very narrow regime of the parameter space. 
We believe that attainment of the self-consistency is the most crucial issue in all these studies, owing to, possibly a large number of competing ground states. 
We have been able to address this difficulty partially by a very thorough numerical investigation of the problem.

Organization of the paper is as follows: a brief introduction to the model Hamiltonian is given in Section-II. 
In Section-III, we calculate the critical exchange potential, $J_{sc}$ for two particles in 2D using the $t$-$J$-$U$ model. 
In Section-IV, we consider finite particle density and investigate superconducting pairing in general, and the impact of on-site Coulomb repulsion on 
the pairing correlations, and finally on BCS-BEC crossover by calculating the chemical potential and mean-pair radius. Finally, we conclude in Section-V.

\section{Model Hamiltonian}
We model the 2D $d$-wave SC by the Hamiltonian,
\begin{eqnarray}  
{\cal H} &=&{\cal H}_{\rm {KE}} + {\cal H}_{\rm int} + {\cal H}_{\rm {\mu}} \nonumber \\
         &=&-t\sum_{\langle{ij}\rangle,\alpha} (c_{i\alpha}^{\dag} c_{j\alpha} + h.c.) 
              + J\sum_{\langle{ij}\rangle}\left({\bf S}_i \cdot {\bf S}_j - \frac{n_i.n_j}{4} \right) \nonumber \\ 
              &&+ U \sum_{i} n_{i \uparrow} n_{i \downarrow} - \mu \sum_i n_i~.  
\label{Ham}        
\end{eqnarray}
The first term describes the kinetic energy, where 
$t$ is the hopping amplitude of electrons between nearest-neighbors $\langle{ij}\rangle$ with spin $\alpha$,   
$J$ is the pairing interaction strength between the nearest neighbour sites and $U$ is the 
on-site repulsive interaction. It is easy to show that the interaction terms lead to a $d$-wave SC ground state 
in the singlet channel.\cite{kagan}
The spin operators are defined by $S^a_i = c_{i\alpha}^{\dag}\sigma^a_{\alpha\beta}c_{i\beta}$,
where the $\sigma^a$ are Pauli matrices, and the density operators
$n_{i\alpha}= c_{i\alpha}^{\dag}c_{i\alpha}$ with $n_i = n_{i \uparrow} + n_{i \downarrow}$.
The chemical potential $\mu$ is adjusted to fix the average carrier density 
$n = \frac{1}{N}\sum_i \langle n_i \rangle$. 
%
\section{Two Particle Bound State}
For the sake of completeness and to compare and contrast with the results for finite electron density, we present the 
calculation of threshold exchange interaction in the extreme dilute limit, that is for two electrons in an empty lattice. 
This specifies the condition for a two-particle $d$-wave bound state.

\begin{figure}[t]
\begin{center}
 \includegraphics[width=90mm]{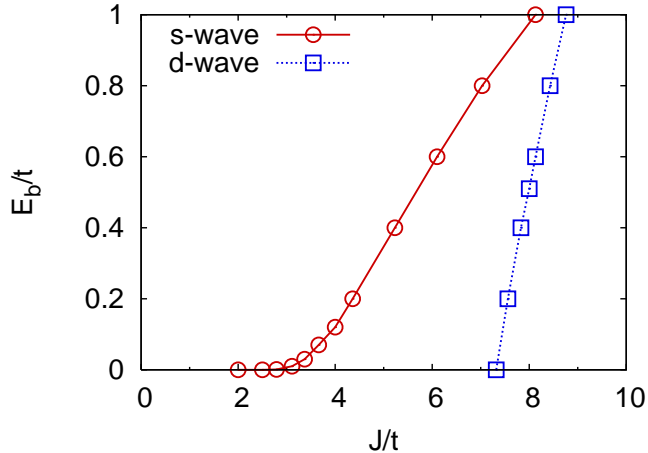}
\caption{(color online) Behavior of binding energy $E_{b}/t$ as function of pairing strength $J/t$ for a isotropic square lattice with  
$s$-wave and $d$-wave symmetry. The critical pairing strengths for bound state formation are $2t$ and $7.32t$ in case of s- and $d$-wave, respectively.}\label{sd}
\end{center}
\end{figure}

Here we start with the particle wave function that is appropriate for a singlet pairing,
\begin{eqnarray}\centering
|\Psi\rangle=\sum_{i_{1},i_{2}}\Phi(i_{1},i_{2})c_{i_{1}\uparrow}^{\dagger}c_{i{2}\downarrow}^{\dagger}|0\rangle~,
\end{eqnarray} 
where $\Phi(i_{1},i_{2})=\Phi(i_{2},i_{1})$. 
Thus the equation of motion,
 \begin{math}{\cal H}|\Psi\rangle={\cal E}|\Psi\rangle\end{math} can be written as (${\cal E}$ being the energy of the electron pairs),
\begin{eqnarray}
{\cal E}\Phi(i_{1},i_{2})&=& \sum_{j}[t_{i_{1}j}\Phi(j,i_{2})+t_{i_{2}j}\Phi(i_{1},j)]+[U\delta_{i_{1},i_{2}}-J_{i_{1},i_{2}}]\Phi(i_{1},i_{2})
\end{eqnarray}
Fourier transform of the equation yields,
\begin{eqnarray}
{\cal E}\Phi(\mathbf{k_{1},k_{2}}) &=& [t(\mathbf{k_{1}})+t(\mathbf{k_{2}})] \Phi(\mathbf{k_{1},k_{2}})+\frac{U}{N}\sum_{k}\Phi(\mathbf{k_{1}+k,k_{2}-k}) \nonumber \\
&& - \frac{1}{N}\sum_{k}J(\mathrm{k})\Phi(\mathbf{k_{1}-k,k_{2}+k})~,
\end{eqnarray}
where 
\begin{equation}
\Phi(\mathbf{k_{1},k_{2}})=\frac{1}{N}\sum_{i_{1},i_{2}}\Phi(i_{1},i_{2})\times
\exp[{-i\lbrace\mathbf{k_{1}}\cdot\mathbf{r_{i_{1}}}+\mathbf{k_{2}}\cdot\mathbf{r_{i_{2}}}\rbrace}]~, \nonumber       
\end{equation}
$t(\mathbf{k})=-2t(\cos{k_{x}}+\cos{k_{y}})$ and
$J(\mathbf{k})=2J(\cos{k_{x}}+\cos{k_{y}})$, where the lattice constant is chosen to be unity.
Let us define $\mathbf{Q}=\mathbf{k_{1}}+\mathbf{k_{2}}, \mathbf{q}=\frac{1}{2}(\mathbf{k_{1}}-\mathbf{k_{2}})$, 
and $\Phi(\mathbf{k_{1}},\mathbf{k_{2}})=\Phi_{Q}(\mathbf{q})$ then we obtain
\begin{equation}
\Phi_{Q}(\mathbf{q})=\frac{\frac{U}{N}\sum_{k}\Phi_{Q}(\mathbf{k})-\frac{1}{N}\sum_{k}J(\mathbf{q}-\mathbf{k})\Phi_{Q}(\mathbf{k})}
{E-t(\frac{\mathbf{Q}}{2}+\mathbf{q})-t(\frac{\mathbf{Q}}{2}-\mathbf{q})}~.\label{phiq}
\end{equation}
For a singlet pairing, we can take Q = 0 and Eq. (\ref{phiq}) can be decoupled so that we can write
\begin{eqnarray}
C_{0} &=& UI_{0}C_{0}-2JI_{x}C_{x}-2JI_{y}C_{y} \nonumber \\
C_{x} &=& UC_{0}I_{x}-2JC_{x}I_{xx}-2JC_{y}I_{xy} \nonumber \\
C_{y} &=& UC_{0}I_{y}-2JC_{x}I_{xy}-2JC_{y}I_{yy}~.\label{ci}
\end{eqnarray}
$C$'s and $I$'s are lattice integrals which can be defined as,
\begin{eqnarray}
C_{0} &=& \frac{1}{N}\sum_{\mathbf{q}}\Phi_{0}(\mathbf{q})~, \qquad
C_{l} = \frac{1}{N}\sum_{k}\cos{k_{l}}\Phi_{0}(\mathbf{k})~, \nonumber \\
I_{0} &=& \frac{1}{N}\sum_{q}{\frac{1}{E+4t(\cos{q_{x}+\cos{q_{y}})}}}~, I_{xy} = \frac{1}{N}\sum_{q}{\frac{\cos{q_{x}}\cos{q_{y}}}{E+4t(\cos{q_{x}}+\cos{q_{y}})}}~,\nonumber \\
I_{l} &=& \frac{1}{N}\sum_{q}{\frac{\cos{q_{l}}}{E+4t(\cos{q_{x}+\cos{q_{y}})}}}~, I_{ll} = \frac{1}{N}\sum_{q}{\frac{\cos^{2}{q_{l}}}{E+4t\cos{q_{l}}}},\nonumber \\
\end{eqnarray}
where $l\in\lbrace x,y\rbrace$.
For an isotropic square lattice symmetry $I_{y}=I_{x}$, $I_{yy}=I_{xx}$ and $C_{y}=C_{x}$.
Unique solutions of $C_{0}$, $C_{x}$ and $C_{y}$ can be obtained iff the determinant of the coefficient
matrix is zero i.e,
\begin{equation}
\left |\begin{array}{ccc}
UI_{0}-1 & -2JI_{x} & -2JI_{x}\\
UI_{x} & -2JI_{xx}-1 & -2JI_{xy}\\
UI_{x} & -2JI_{xy} & -2JI_{xx}-1
\end{array}\right|=0~.\label{det}
\end{equation}
To project out the possibility of double occupancy we now consider $U\rightarrow\infty$. 
This reduces the Hamiltonian in Eq. (\ref{Ham}) to a more familiar variant, the vastly studied $t$-$J$ model. 
It is worth mentioning that the lattice integrals can be evaluated analytically and one can verify that only 
two integrals are completely independent, such that we only need
\begin{eqnarray}
I_{0} &=& \frac{1}{2}\frac{K(\frac{-2}{\alpha})}{\pi\alpha} \nonumber \\
I_{xx} &=& \frac{1}{4}\frac{({\alpha}^2-2\alpha+2)K(\frac{-2}{\alpha})}{\pi\alpha}-\frac{1}{2}\frac{(\alpha-2)\Pi(\frac{2}{\alpha},
\frac{-2}{\alpha})}{\pi}+\frac{1}{4}\frac{{\alpha}E(-\frac{2}{\alpha})}{\pi}~,\label{elliptic}
\end{eqnarray}
where $\alpha$ is a dimensionless energy, $\frac{\cal E}{4t}$. $K$, $E$ and $\Pi$ denotes the complete elliptic integral
of first, second and third kind respectively.
The rest of the integrals can be represented using these two in the following way,
\begin{eqnarray}
I_{xx}+I_{xy} &=& -\frac{E}{4t}I_{x}, \qquad\textrm{and}\qquad I_{x} = \frac{1}{8t}-\frac{E}{8t}I_{0}~.\label{IxI0}
\end{eqnarray}
So the lattice integrals can be computed in terms of complete elliptic integral 
of the first kind $K(-2/\alpha)$, the second kind $E(-2/\alpha)$, and the third kind $\Pi(2/\alpha,-2/\alpha)$ 
respectively \cite{Jer}. 
The critical value of $J$ can be obtained by substituting ${\cal E}=-8t-\delta$, i.e the enegy
for the electron-pair lying just below the noninteracting two electron band in 2D ($=-8t$) 
from Eqs. (\ref{elliptic}, \ref{IxI0}).
Expansion of the elliptic integrals for $\delta\rightarrow0$ yields a
logarithmic divergence. However the coefficient of the diverging term should be made to vanish, which leads to a quadratic equation for $J$, viz.
\begin{eqnarray}
 (4-\pi)J^2-8J+4\pi=0~.\label{bs}
\end{eqnarray}
The solutions of Eq. (\ref{bs}) are as $J_c$ = $2t$ and $7.32t$, which actually leads to the critical value of 
bound state formation for two electrons corresponding to $s$-wave and $d$-wave symmetries. Conversely, from Eq. (\ref{det})
we obtain a pair of equations as,
\begin{eqnarray}
 J^{s}&=&-\frac{1-UI_{0}}{2(2UI_{x}^2+(I_{xx}+I_{xy})(1-UI_{0}))}\nonumber\\
J^{d}&=&-\frac{1}{2(I_{xx}-I_{xy})}~.\label{IsId}
\end{eqnarray}
The first equation in Eq. (\ref{IsId}) corresponds to the $s$-wave case and the second one to the $d$-wave. One should note that
the $d$-wave bound state is independent of $U$. 
The $s$- and $d$-wave bound state energies can
be written as,
\begin{eqnarray}
 E_{b}^{s}&=&-8t-64t\exp{\Big[-\frac{\pi J}{J-J_{c}^{s}}\Big]}~,\nonumber\\
 E_{b}^{d}&=&-8t-\frac{16t^2}{1-\frac{2}{\pi}}\Big[-\frac{1}{J_{c}^{d}}-\frac{1}{J}\Big]~,
\end{eqnarray}
where $J_{c}^{s,d}$ being the critical value of $J$ in $s$- and $d$-wave cases. 
In Fig. (\ref{sd}), we show the dependence of the bound state energies for the $s$- and $d$-wave cases 
as a function of the exchange interaction, $J$.
It can be seen that the $s$-wave critical value drops exponentially at lower $J$ values, whereas the $d$-wave curve has a near-linear fall off. 
However a $s$-wave bound state is energetically 
more favorable compared to the other. It is interesting to note that, $J_{c}^{d}$ is higher that $J_{c}^{s}$, despite
the fact the latter is affected by the strong on-site repulsion. 
Thus we conclude that for $J>J_{c}^{s}$ at low density, the ground state of the system is a Bose condensate of fermion pairs having a $s$-wave 
pairing symmetry.

Hence we analyze pairing correlation in moderately high particle density regime in the coming section, 
where a crossover to a BEC state is further investigated.
Competition between the two interactions results in 
(a) a rich behavior of $J_{sc}$ with increasing $n$ and $U$, and 
(b) BCS-BEC crossover close to half filling, in contrast to the existing results.

\section{BCS-BEC crossover in a $d$-wave superconductor}

To solve the many-particle Hamiltonian (Eq. \ref{Ham}) we reduce the quartic terms to quadratic terms with a mean-field decoupling
and then employ Bogoliubov transformation 
\begin{equation}
 c_{i\alpha} = \sum_n \left(u_{in\alpha} \gamma_{n\alpha}^{} -\alpha v^{*}_{in\alpha} \gamma_{n\overline{\alpha}}^{\dagger} \right)
\end{equation}
to diagonalize the mean-field Hamiltonian, which takes the form
\begin{equation}
H = \sum_{n\alpha} E_{n\alpha} \gamma_{n\alpha}^{\dagger} \gamma_{n\alpha}^{}~.
\end{equation} 
$\gamma_{n,\alpha}^\dag$  creats an elementary fermionic Bogoliubov 
quasiparticle excitation with quantum number $n$, spin $\alpha$, and energy $E_{n\alpha}>0$. 
Calculation of the commutators of the above $H$ with the electron operators $c_{i\sigma}$ leads 
to the BdG equations
\begin{equation}
\left(\matrix{\hat\xi & \hat\Delta \cr \hat\Delta^{*} & -\hat\xi^{*}} \right)
\left(\matrix{u_{n} \cr v_{n}} \right) = E_{n}
\left(\matrix{u_{n} \cr v_{n}} \right)
\label {eq:bdg}
\end{equation}
where
$\hat\xi u_{n}(i) = -\sum_{\delta}(t + W_i) u_{n}(i+\delta)  
-\tilde{\mu}_iu_{n}(i)$ and 
$\hat\Delta u_{n}(i) = \sum_{\delta}\Delta(i,i+\delta) u_{n}(i+\delta)
$,
and similarly for $v_{n}(i)$. 
The pairing amplitudes on a bond is defined by
$\Delta(i;i+\delta) = -J\langle c_{i+\delta \downarrow}c_{i \uparrow}
+ c_{i \downarrow}c_{i+\delta \uparrow}\rangle/2$, where 
$\delta = \pm{\hat{x}}, \pm{\hat{y}}$.
The Hartee-Fock shifts are given by
$\tilde{\mu}_i = \mu - U \langle n_{i} \rangle
+ \frac{J}{2} \sum_{\delta}\langle n_{i+\delta} \rangle$
and $W_i = \frac{J}{2} \langle c_{i+\delta,-\alpha}^{\dag}c_{i, \overline{\alpha}} \rangle$. 
Starting with an initial guess for all local variables defined on the sites and bonds of the 
lattice, we numerically solve for the BdG eigenvalues for $E_n \ge 0$ and the corresponding eigenvectors 
$\left(u_{n},v_{n}\right)$. The pairing amplitude is given by  
\begin{eqnarray}
\Delta(i;i+\delta) = \frac{J}{2}\sum_n\left[u_n(i+\delta)v^*_n(i) 
+ u_n(i)v_n^*(i+\delta)\right]
\end{eqnarray}
with the density
$\langle n_i \rangle = 2\sum_n |v_n(i)|^2$, and the Fock shift
$W_i = \frac{J}{2}\sum_n v_n(i+\delta)v^*_n(i)$ are calculated iteratively until self-consistency is achieved. 
The $d$-wave pairing amplitude is defined as,
\begin{eqnarray}  
\Delta_d(i) = \left[\Delta(i;+\hat{x})-\Delta(i;+\hat{y})
+ \Delta(i;-\hat{x}) - \Delta(i;-\hat{y}) \right]/4~, \nonumber 
\end{eqnarray}
where $\hat{x}$ and $\hat{y}$ connect to the nearest neighbors in 2D.

It is worthwhile to mention at this point that achieving self-consistency in our calculations is a significantly crucial issue, as a number of parameters demand 
simultaneous convergence, which as experienced by us, is a challenging task possibly because of a number of competing ground states. 
May be due to this difficulty Ghosal $et.$ $al.$ have done their calculations by taking only one set of parameter values.\cite{ghosal} 
Thus an extensive search for self-consistent solution over a broad parameter space has been done, however with limited success. Yet in our work we were able to scan a reasonable amount of parameter space defined by $U$, $J$ and $n$, which has not been done earlier.

We begin by studying the ground state properties and calculate the critical value of pairing energy $J_{sc}$ which we defined as a threshold value 
below which pairing correlations are zero (normal state).  We find that $J_{sc}$ depends strongly on the density and on-site 
Coulomb repulsion. Then, we investigate the impact of Coulomb repulsion on the BCS-BEC crossover picture. 
The density range we are interested in is between quarter and half filling as the situation for lower densities is addressed in earlier studies. 
We consider a $24\times 24$ system size for all our calcuations.  

\begin{figure}[t]
\begin{center}
\includegraphics[width=90mm]{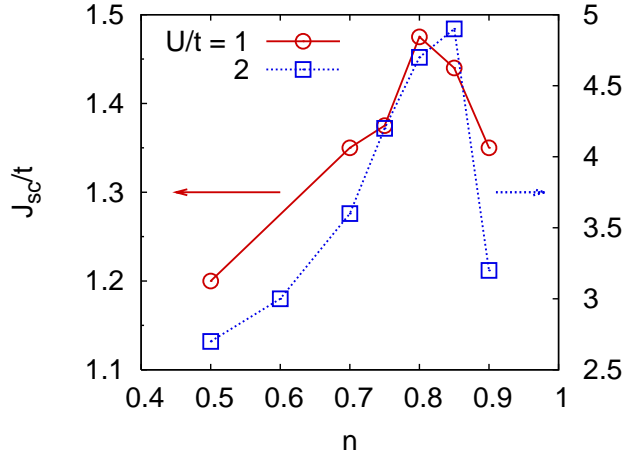}
\caption{(color online) Variation of critical value of pairing strength $J_{sc}/t$ with density for fixed values of $U$ (= 1 and 2). 
$J_{sc}$ shows an optimization behavior with density due to the on-site Coulomb repulsion. The system size considered is $24\times 24$.}
\label{fig:Jmin}
\end{center}
\end{figure}

\begin{figure}[t]
\includegraphics[width=70mm]{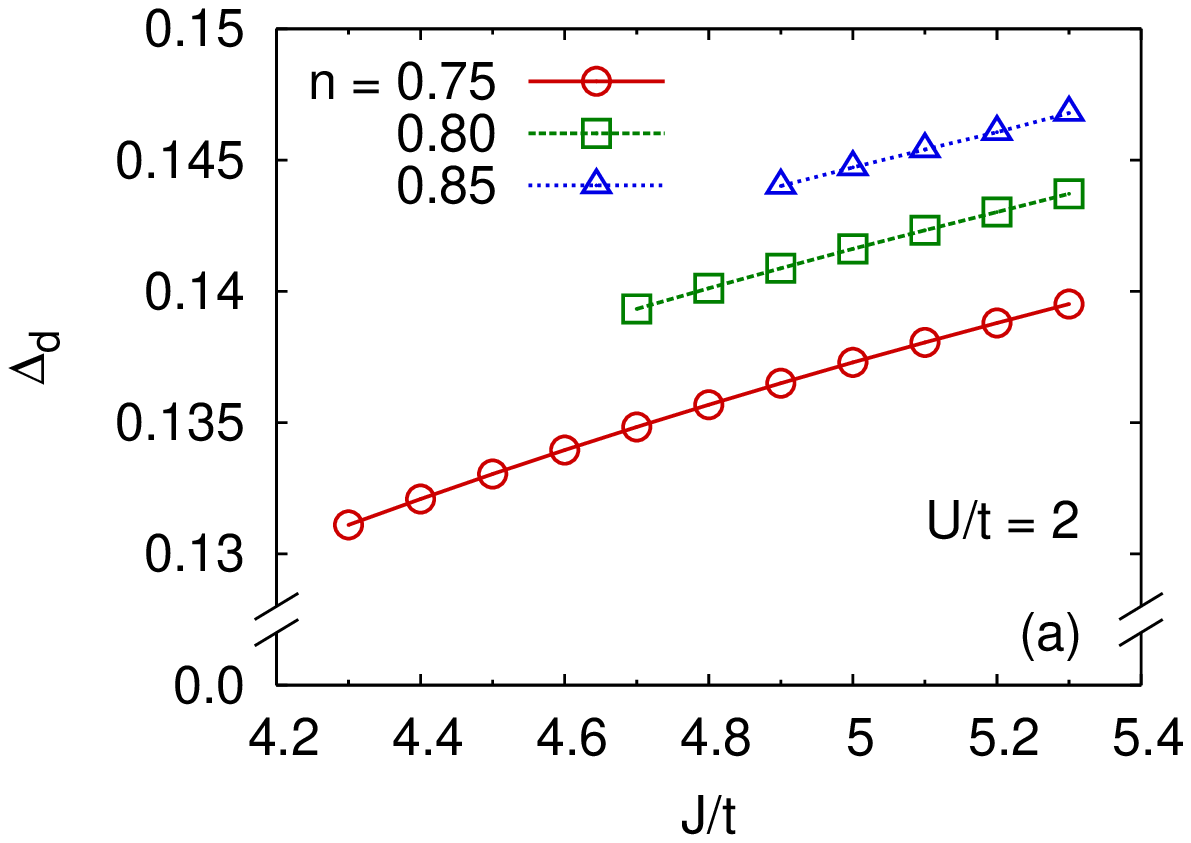}
\includegraphics[width=70mm]{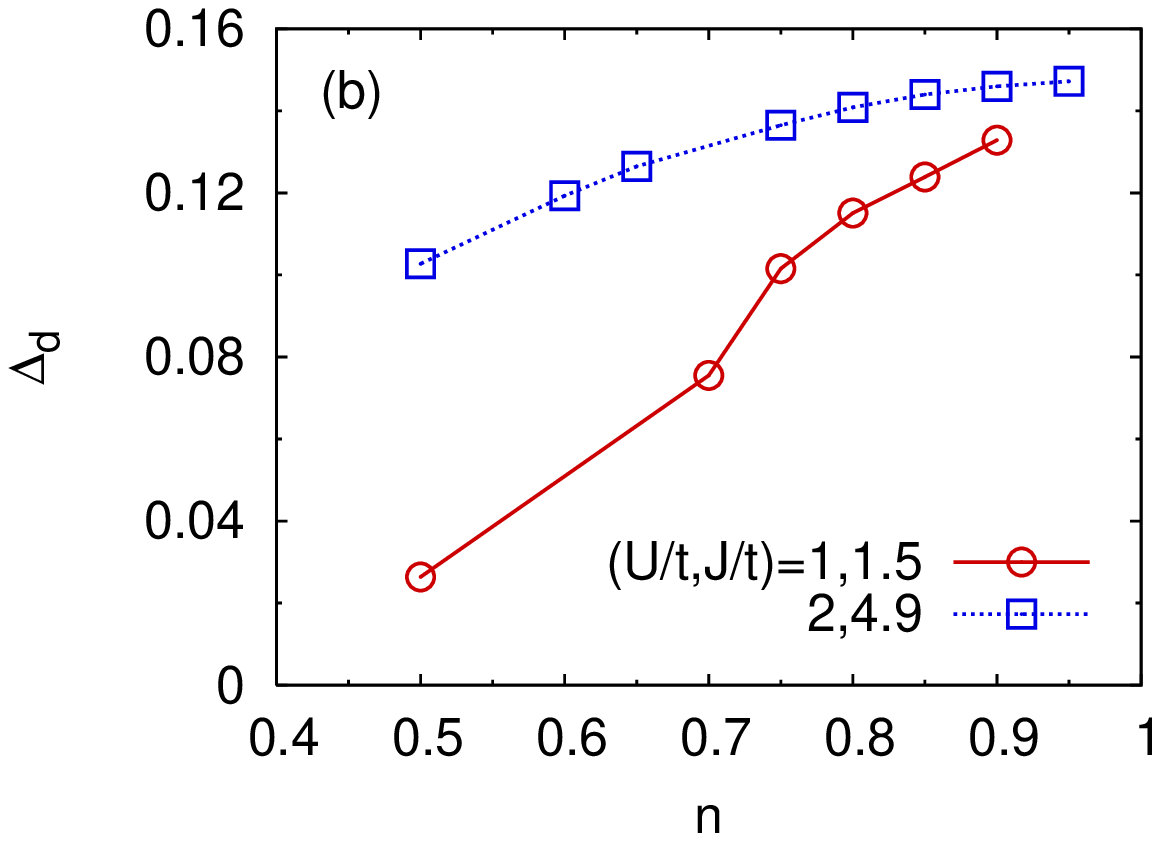}
\caption{(color online) Behavior of $d$-wave correlations $\Delta_d$ (a) with pairing energy $J$ shows a linear grow and (b) with electron density $n$ saturates near half filling ($n$=1) due to complete overlap of the electronic states.
The parameters are taken from Fig. \ref{fig:Jmin} and chosen so that the system always in BCS or BEC state.}
\label{fig:del}
\end{figure}

We have seen in the last section that the threshold for a single bound state of two electrons on the empty lattice of a 2D $t$-$J$ model is $J_c^d \simeq 7.32t$ for
$d$-wave symmetry. With increasing particle density the average spacing between the particles decreases and consequently the overlap of the 
single particle states increases. Thus, the minimum pairing potential required for bound state formation monotonically decreases with density.\cite{hertog} 
In a $t$-$J$-$U$ model, as we increase density, the on-site Coulomb repulsion becomes more effective, which makes it energetically costlier for the electrons to come closer and form pairs. So in our case $J_{sc}$ initially increases with density, with its value always less than 7.32$t$ (Fig. \ref{fig:Jmin}). Close to half filling, the electrons naturally sit next to each other resulting in lower minimum pairing energy. Hence the critical coupling shows a characteristic optimization behavior with particle density.  This non-monotonic behavior stimulate our interest to investigate the BCS-BEC crossover near half filling, where it always shows 
characteristics of a BCS state in the absence of $U$. It may be noted that even though the two-body bound is insensitive to the value of the Coulomb repulsion, 
the pairing correlation in a system with finite carrier density, this is no longer true. Further, as our main focus is the larger density regime, so we do not address densities lower than qurter filling.\cite{hertog} 

Both the attractive pairing potential and electron density drive $d$-wave correlations, $\Delta_d$. 
As $\Delta_d$ is direcly proportional to the pairing potential, it grows linearly with the strength of the potential for a fixed particle density 
(Fig. \ref{fig:del}(a)). Further, the $d$-wave correlations depend upon the overlap of the electronic states, so $\Delta_d$ increases with particle density. Thus, for fixed values of $J$ and $U$ the correlations monotonically increase with density. Close to half filling, the particles start sitting next to each other and consequently the overlap becomes complete, resulting in a saturation behaviour of $\Delta_d$ with density (Fig. \ref{fig:del}(b)).

\begin{figure}
\includegraphics[width=70mm]{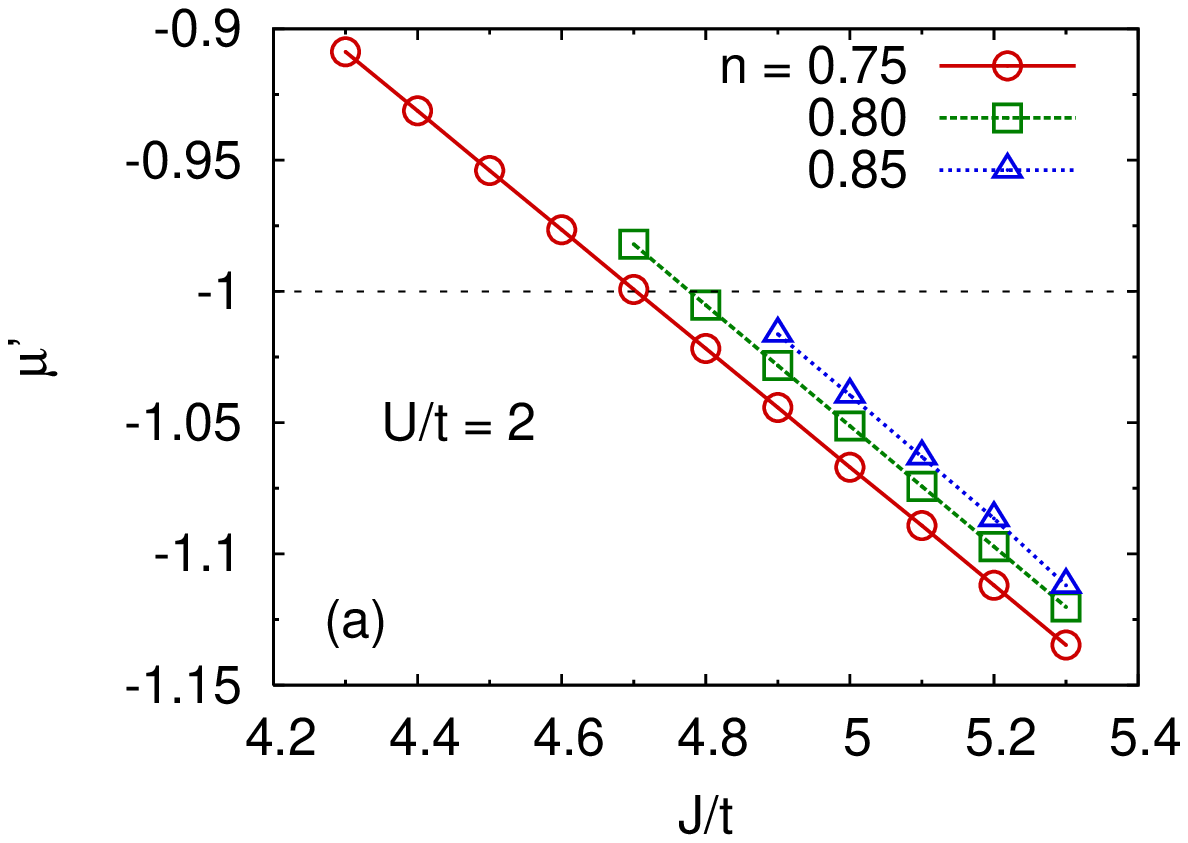}
\includegraphics[width=70mm]{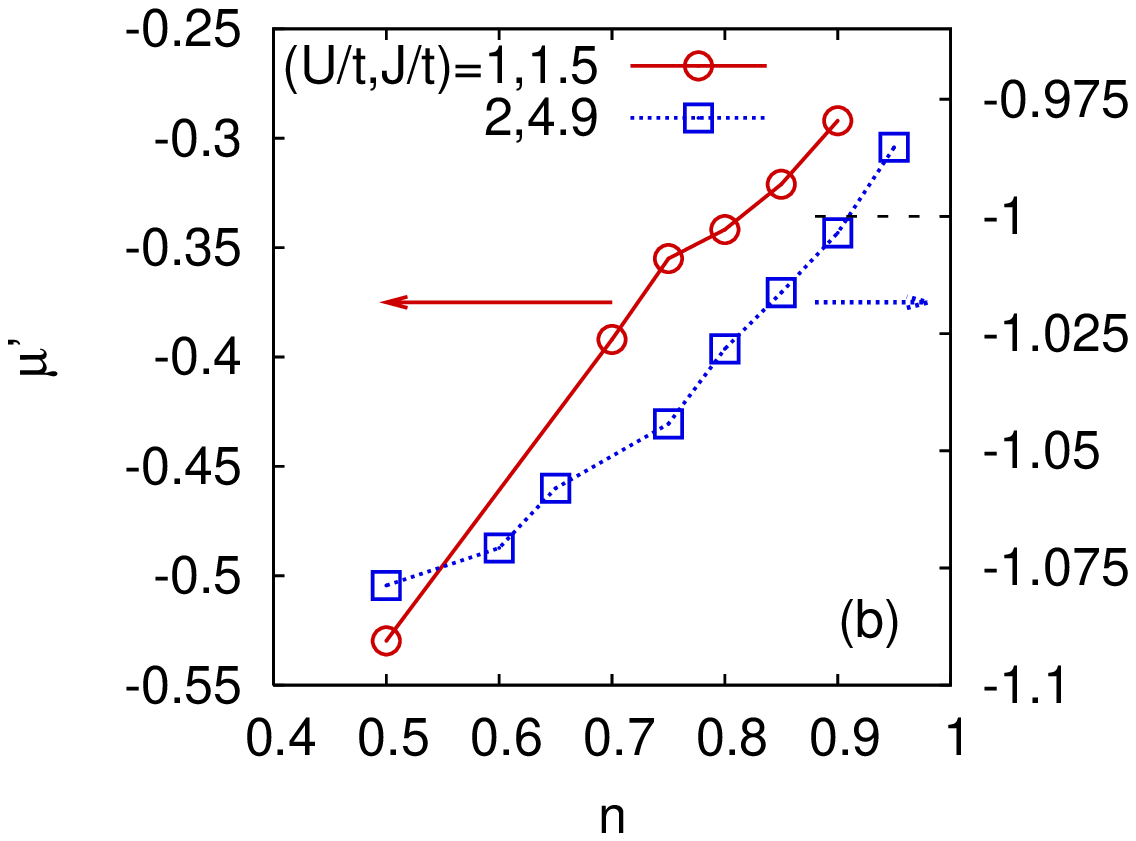}
\caption{(color online) Variation of (scaled) chemical potential, $\mu'$ as a function of (a) pairing energy $J$ and (b) electron density $n$. 
BCS-BEC crossover occures near half filling ($n$=1) for $U$=2 and moderate pairing strength.}
\label{fig:mu}
\end{figure}

The BCS-BEC crossover scenario is conveniently investigated by calculating the chemical potential 
as a function of the interaction strength between the fermions\cite{leggett,tifrea}. 
In general, bosonic degrees of freedom is expected to emerge once the chemical potential of the many-body ground state 
slips below the noninteracting band minimum in a tight-binding system.  
In a $s$-wave system, a smooth crossover from fermionic superconductivity to bosonic degree of freedom can occur 
for all densities as the coupling strength is increased.\cite{hertog}  
In a $d$-wave system, without the on-site repulsion bosonic degrees of freedom can only emerge in the extreme dilute limit, while 
for the large densities, the system behaves more like a weak-coupling superconductor\cite{hertog}. 
Although, as mentioned erlier, the on-site Coulomb repulsion is not indispensably necessary for generating a two-body
$d$-wave pairing state, we find that it has a significant role in $d$-wave pair formation and the BCS-BEC crossover phenomena.

Fig. \ref{fig:mu}(a) shows the crossover phenomena for moderately high densities with (stronger) pairing strength $J$. Here we have set $U/t$=2. 
The scaled chemical potential $\mu'$ = $\mu$/4t slips below the noninteracting band for $n$ = 0.75 and 0.8, in contrast to the existing results 
where BCS-BEC crossover occures for low densities only.\cite{hertog} Interestingly, for a higher $n$ (= 0.85), the system directly goes from normal 
to a BEC state without visiting the conventional BCS state.

Next, we explore the BCS-BEC crossover with variation of $n$ for fixed values of $U$ and $J$. 
For moderately weak parameter values, $J/t=1.5$ and $U/t=1$, the system prefers to stay in the BCS state for the calculated densities (Fig.\ref{fig:mu}(b)), consistent with the existing results. Interstingly, for the same density values, the system goes to the BEC state as we increase the pairing strength and Coulomb repulsion, $J/t=4.9$ and $U/t=2$. Further, the system goes to a BCS state quite close to half filling. These results are new and indicate the importance of Coulomb repulsion on the crossover mechanism.  
The reason for choosing unusual parameter values ($e.g.$ $J/t$=1.5 and 4.9) is that, inspite of a thorough and careful search for the self-consistent solutions 
at other parameter values, the success was limited. However we could scan a broad density regime which has contributed to a lot more enriched study of the problem compared to what already exists.

\begin{figure}
\begin{center}
\includegraphics[width=90mm]{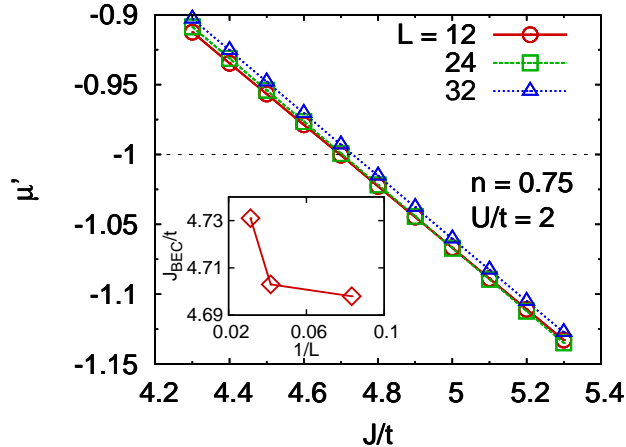}
\caption{(color online) Variation of (scaled) chemical potential, $\mu'$ as a function of pairing energy $J$ for different system sizes $L$ = 12, 24, and 32 with $n$ = 0.75 and $U/t$ = 2 fixed. Also shown $J_{BEC}/t$ with 1/$L$, which changes only in the second decimal place as system size is increased.}
\label{fig:scaling}
\end{center}
\end{figure}

To investigate the finite-size scaling effects of the the crossover scenario, we have considered lattices of different sizes, namely $L$ = 12, 24, and 32 for fixed value of density $n$ = 0.75 and Coulomb repulsion $U/t$ = 2. As shown in Fig. \ref{fig:scaling}, the crossover phenomena is qualitatively unaffected as system size is increased. In the inset we have plotted $J_{BEC}/t$ with 1/$L$, where $J_{BEC}$ is the critical value of the pairing strength at which the BCS-BEC crossover occurs. It is clear that the crossover phenomena is minimally affected and changes only in the second decimal place as system size is increased, which will not affect our qualitative study and leave the  notion of a BCS-BEC crossover in a d-wave superconductor intact.

\begin{figure}[t]
\includegraphics[width=70mm]{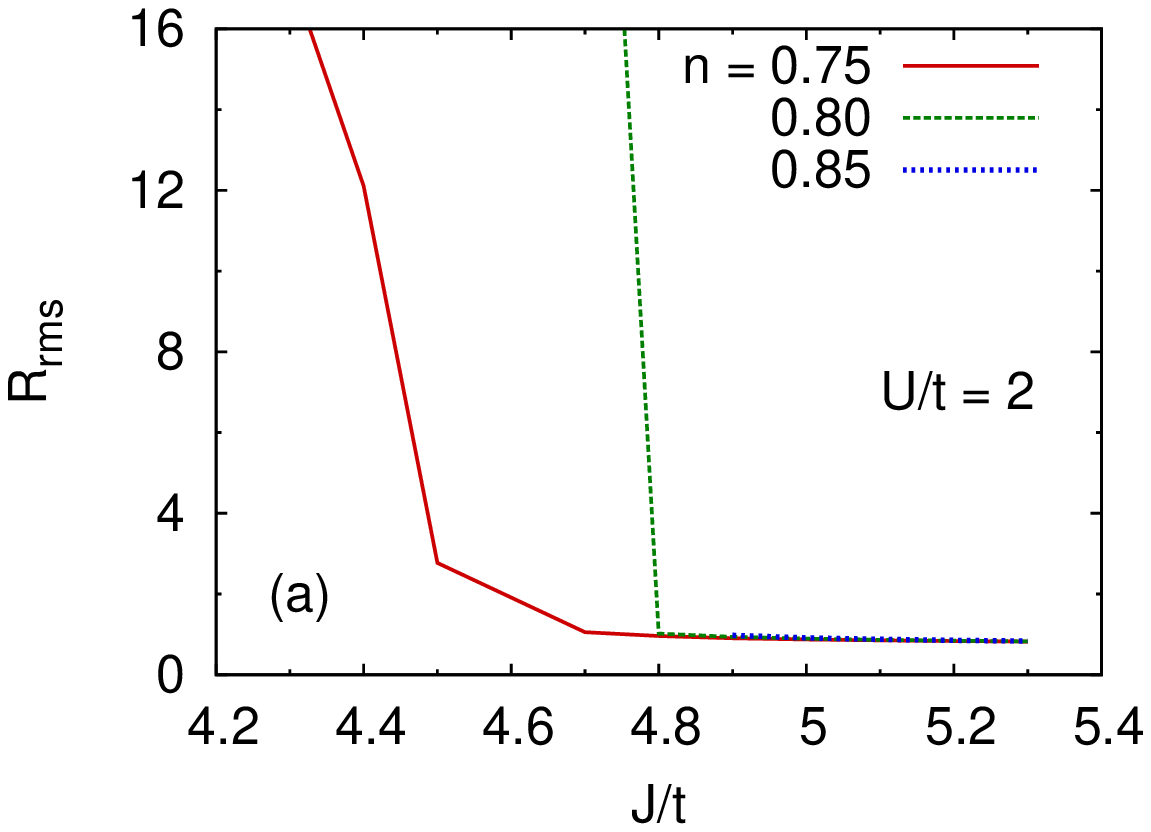}
\includegraphics[width=70mm]{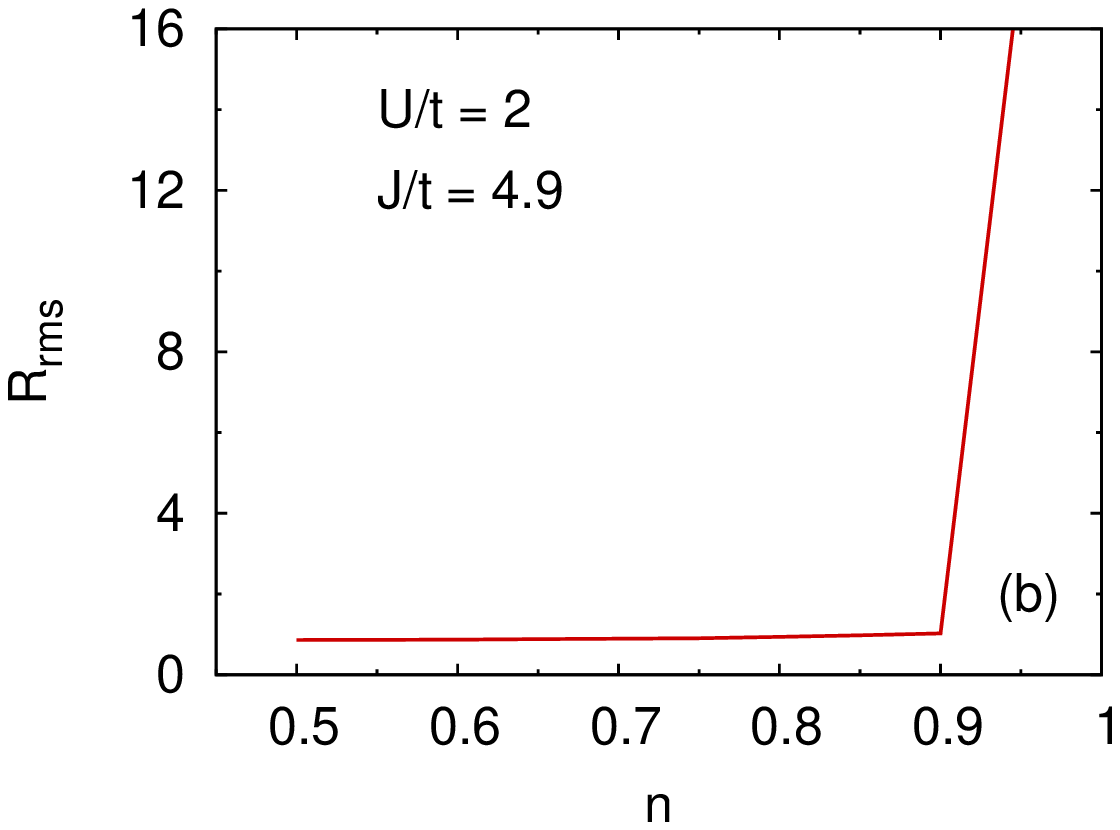}
\caption{(color online) Variation of root mean squre radius $R_{rms}$ as a function of (a) pairing energy $J$ and (b) electron density $n$. 
$R_{rms}$ shrinks from a few tens of lattice spacing to order one of lattice spacing across BCS-BEC crossover and consistent with the chemical potential 
picture (Fig. \ref{fig:mu}).}
\label{fig:rad}
\end{figure}

In case of the $s$-wave superconductors, largely overlapping of Cooper pairs (BCS) smoothly evolve into short ranged tightly bound pairs (BEC) 
with increasing interparticle attraction among the fermions.
In $d$-wave systems, the pairs always retain a finite spatial extent. A better physical picture of the BCS-BEC crossover 
can be given by calculating the average pair coherence length or the mean pair radius $R_{rms}$.
It is defined by the following relation\cite{song},
\begin{equation}
R_{rms}^2 = \frac{\int\mid f(r)\mid^2r^2d^3r}{\int\mid f(r)\mid^2d^3r} = \frac{\sum\mid\nabla_kg_k\mid^2}{\sum\mid\ g_k\mid^2}~,
\end{equation}
where $f(r)$($g(k)$) is the wave function for a Cooper pair in real (momentum) space. 

Fig. \ref{fig:rad}(a) shows the evolution of $R_{rms}$ as a function of the exchange potential and at a fixed density in a 24 $\times$ 24 lattice. 
Clearly, $R_{rms}$ shrinks from a few tens of lattice spacing to order one of lattice spacing as the attracting pair strength increases. 
The transition takes place precisely at the same value of $J$ where $\mu'$ slips below the noninteracting band edge.
This supports the proposed crossover scenario in a $d$-wave superconductor. 
For a fixed value of exchange potential, the overlap of the electronic states increases with particle density, which supresses 
the emergence of bosonic degrees of freedom near half filling and the system instead remains fermionic (BCS-like), even for moderately stronger couplings. 
This is manifested in Fig. \ref{fig:rad}(b), where for $n$ $>$ 0.9 the system shows characteristics of a BCS superconductor.  
In our study, the mean pair radius and the chemical-potential descriptions of BCS-BEC crossover are fully consistent with each other 
(Fig. \ref{fig:mu} and Fig. \ref{fig:rad}). 
In contrast to the earlier results, where the crossover is seen to occur only in the dilute limit,\cite{hertog} we have seen that it can be realized near half filling.

\section{Conclusions}
The BCS-BEC crossover has been investigated for $d$-wave superconductors in 2D using a $t$-$J$-$U$ model near half filling. We have employed the mean-field Bogoliubov-de Gennes method and calculated the ground state properties self consistently. To compare and contrast between the zero and large density limits, we included a discussion on the threshold exchange potentials that are required to form a two particle bound state (at zero density) and superconductivity (at large densities). While the former has no effect on the Coulomb repulsion, $U$, the latter seems to have been affected rather significantly by it. Further, the BCS-BEC crossover scenario is investigated as a function of both the exchange interaction and Coulomb repulsion, and the possibility of the crossover is concluded for some 
specific parameter values. To complement the earlier studies on the subject at low densities, the crossover picture is demonstrated at high densities by studying the 
behavior of the chemical potential which when falls below the noninteracting band minimum, signals the onset of a BEC like phase. A robust support of the crossover 
is provided by computing the pair radius which shrinks from tens of lattice spacing to that of a very few at the onset of crossover. 

It is worthwhile to mention that going beyond the standard mean-field approach by including pair fluctuations within the attractive Hubbard model, it was found that around half filling the smooth evolution from the BCS to the BEC limits is interrupted.\cite{chien} The transition temperature to a superfluid phase $T_c$ vanishes near half filling over an extended range for moderately strong attraction when the system approaches the bosonic regime. This vanishing is associated with the divergence of the mass of the pairs or localization of the pairs due to pairing fluctuations, which eventually destroys the superfluid state. Also, as the density approaches half filling the chemical potential get pinned to its noninteracting value due to particle-hole symmetry. So, although $T_c$ vanishes due to strong interaction, the system still stays in the fermionic regime ($\mu > 0$). Thus including fluctuation effects to study the crossover phenomena may be a worthwhile exercise for the future.

\end{document}